\documentclass[aps,prd,twocolumn,superscriptaddress,showpacs]{revtex4}
\usepackage[dvipdfmx]{graphicx}
\usepackage{amsmath,amssymb,times}
\usepackage{color}

\newcommand{\bequ}{\begin{equation}}
\newcommand{\eequ}{\end{equation}}
\newcommand{\bea}{\begin{eqnarray}}
\newcommand{\eea}{\end{eqnarray}}

\DeclareSymbolFont{boldletters}{OML}{cmm} {b}{it}
\DeclareSymbolFontAlphabet{\mathbit}{boldletters}
\DeclareMathSymbol{\alpha}{\mathalpha}{letters}{"0B}
\DeclareMathSymbol{\beta}{\mathalpha}{letters}{"0C}
\DeclareMathSymbol{\gamma}{\mathalpha}{letters}{"0D}
\DeclareMathSymbol{\delta}{\mathalpha}{letters}{"0E}
\DeclareMathSymbol{\epsilon}{\mathalpha}{letters}{"0F}
\DeclareMathSymbol{\zeta}{\mathalpha}{letters}{"10}
\DeclareMathSymbol{\eta}{\mathalpha}{letters}{"11}
\DeclareMathSymbol{\theta}{\mathalpha}{letters}{"12}
\DeclareMathSymbol{\iota}{\mathalpha}{letters}{"13}
\DeclareMathSymbol{\kappa}{\mathalpha}{letters}{"14}
\DeclareMathSymbol{\lambda}{\mathalpha}{letters}{"15}
\DeclareMathSymbol{\mu}{\mathalpha}{letters}{"16}
\DeclareMathSymbol{\nu}{\mathalpha}{letters}{"17}
\DeclareMathSymbol{\xi}{\mathalpha}{letters}{"18}
\DeclareMathSymbol{\pi}{\mathalpha}{letters}{"19}
\DeclareMathSymbol{\rho}{\mathalpha}{letters}{"1A}
\DeclareMathSymbol{\sigma}{\mathalpha}{letters}{"1B}
\DeclareMathSymbol{\tau}{\mathalpha}{letters}{"1C}
\DeclareMathSymbol{\upsilon}{\mathalpha}{letters}{"1D}
\DeclareMathSymbol{\phi}{\mathalpha}{letters}{"1E}
\DeclareMathSymbol{\chi}{\mathalpha}{letters}{"1F}
\DeclareMathSymbol{\psi}{\mathalpha}{letters}{"20}
\DeclareMathSymbol{\omega}{\mathalpha}{letters}{"21}
\DeclareMathSymbol{\varepsilon}{\mathalpha}{letters}{"22}
\DeclareMathSymbol{\vartheta}{\mathalpha}{letters}{"23}
\DeclareMathSymbol{\varpi}{\mathalpha}{letters}{"24}
\DeclareMathSymbol{\varrho}{\mathalpha}{letters}{"25}
\DeclareMathSymbol{\varsigma}{\mathalpha}{letters}{"26}
\DeclareMathSymbol{\varphi}{\mathalpha}{letters}{"27}
\DeclareMathSymbol{\Gamma}{\mathalpha}{letters}{"00}
\DeclareMathSymbol{\Delta}{\mathalpha}{letters}{"01}
\DeclareMathSymbol{\Theta}{\mathalpha}{letters}{"02}
\DeclareMathSymbol{\Lambda}{\mathalpha}{letters}{"03}
\DeclareMathSymbol{\Xi}{\mathalpha}{letters}{"04}
\DeclareMathSymbol{\Pi}{\mathalpha}{letters}{"05}
\DeclareMathSymbol{\Sigma}{\mathalpha}{letters}{"06}
\DeclareMathSymbol{\Upsilon}{\mathalpha}{letters}{"07}
\DeclareMathSymbol{\Phi}{\mathalpha}{letters}{"08}
\DeclareMathSymbol{\Psi}{\mathalpha}{letters}{"09}
\DeclareMathSymbol{\Omega}{\mathalpha}{letters}{"0A}

\begin{document}
\preprint{SAGA-HE-286}

\title{Equation of state and transition temperatures in the quark-hadron hybrid model}

\author{Akihisa Miyahara}
\email[]{miyahara@email.phys.kyushu-u.ac.jp}
\affiliation{Department of Physics, Graduate School of Sciences, Kyushu University,
             Fukuoka 819-0395, Japan}

\author{Yuhei Torigoe}
\email[]{torigoe@email.phys.kyushu-u.ac.jp}
\affiliation{Department of Physics, Graduate School of Sciences, Kyushu University,
             Fukuoka 819-0395, Japan}

\author{Hiroaki Kouno}
\email[]{kounoh@cc.saga-u.ac.jp}
\affiliation{Department of Physics, Saga University,
             Saga 840-8502, Japan}

\author{Masanobu Yahiro}
\email[]{yahiro@phys.kyushu-u.ac.jp}
\affiliation{Department of Physics, Graduate School of Sciences, Kyushu University,
             Fukuoka 819-0395, Japan}

\date{\today}

\begin{abstract}
{We analyze the equation of state of 2+1 flavor lattice QCD 
at zero baryon density by 
constructing the simple quark-hadron 
hybrid model that has both quark and hadron components 
simultaneously. We calculate hadron and quark contribution separately and parameterizing those to match with LQCD data. Lattice data on the equation of state are decomposed into 
hadron and quark components by using the model. 
The transition temperature is defined by the temperature at which 
the hadron component is equal to the quark one in the equation of state. 
The transition temperature thus obtained 
is about 215 MeV and somewhat higher than the 
 chiral and the deconfinement pseudocritical temperatures defined by 
the temperature at which the susceptibility or the absolute value of the derivative of the order parameter with respect to temperature becomes maximum. }
\end{abstract}

\maketitle

\section{Introduction}
\label{Introduction}

Lattice QCD (LQCD) has been clarifying properties of the quark-hadron transition at zero quark number chemical potential.  
Two order parameters, the chiral condensate and the Polyakov loop, are commonly used to study the transition. As for the current quark mass $m$, in the limit $m\to 0$, chiral symmetry is exact and the chiral condensate is an order parameter of the spontaneous chiral symmetry breaking. Meanwhile, in the limit $m\to \infty$, 
$Z_3$ symmetry is exact and the Polyakov loop 
defined by 
\begin{eqnarray}
{\rm \Phi}({\bf x})=\frac{1}{N_c}{\rm Tr_c} {\cal P}\exp \left[i\int_0^{1/T} d\tau A_4(\tau,{\bf x})\right], 
\label{polyakov loop}
\end{eqnarray}
is an order parameter of the spontaneous $Z_3$ symmetry breaking, 
where ${\cal P}$ and $A_4 (=iA_0)$ are the path ordering operator and the temporal component of gluon field, respectively, and the trace is taken for the color 
indices. 
Dynamical quark with finite $m$ breaks the $Z_3$ symmetry explicitly.
In the real world with light quarks, the chiral condensate is an approximate but good order parameter of the chiral transition, but it is not clear that the Polyakov-loop is a good approximate order parameter of the 
confinement-deconfinement transition. 
For this reason, in this paper, the transition defined 
by the Polyakov loop is called "$Z_3$ transition" in 
order to distinguish this transition from the so-called 
confinement-deconfinement transition.  
The relation between the chiral restoration and the $Z_3$ transitions is still unclear. 
LQCD simulations with two-flavor dynamical quarks indicate that the 
two transitions take place almost simultaneously at zero baryon density. 
However, in the case of 2+1 flavor dynamical quarks, LQCD simulations show 
that the chiral transition (pseudocritical) temperature $T_c^{C,{\rm L}}$ 
is considerably lower  than the $Z_3$ transition (pseudocritical) 
temperature $T_c^{Z_3,{\rm L}}$~\cite{YAoki_Tc}. 
 It was been argued that nonanalyticity of a certain order parameter propagate 
to the other one~\cite{BCPG,Kashiwa_order}.  
Hence several quantities have nonanalyticity at a common temperature. 
However, at zero density, both the $Z_3$ and chiral transitions are crossover~\cite{YAoki_crossover}, and hence 
the order parameters of these transitions are analytic. 
Therefore, for each of these transitions, the transition temperature is 
commonly defined by 
the temperature at which the susceptibility or the absolute value of the derivative of 
order parameters becomes maximum. 
There is no 
necessity  that the two transitions take place at a common temperature, but 
strong correlations between the two transitions were seen particularly 
in two-flavor LQCD simulations. 
There were several trails to reproduce the correlations. 
For example, in the Polyakov-loop extended Nambu-Jona-Lasinio (PNJL) type models,  
these transitions are correlated
~\cite{MO,Dumitru,Fukushima,Ratti,Megias,Rossner,Schaefer,Abuki,Kashiwa1}.   
However, the original PNJL model predicts a rather higher critical temperature for chiral transition than for $Z_3$ transition,  
if we set the model parameters so as to reproduce 
LQCD data on the $Z_3$ transition temperature. 
In the entanglement PNJL (EPNJL) model~\cite{Sakai_EPNJL,Gatto_EPNJL,Sasaki_EPNJL} and nonlocal PNJL model~\cite{Kashiwa_nonlocal}, there is a strong correlation between the two transitions. 
The two models are successful in reproducing the two-flavor LQCD data  in which the two transitions take place almost simultaneously. However, it is very difficult for the PNJL-type models  to reproduce the 2+1 flavor LQCD data in which the chiral transition temperature is considerably lower than that of the $Z_3$ transition. 
 On the other hand, it is known that, at low temperature, the hadron resonance gas (HRG) model well accounts for LQCD data on the equation of state (EOS) and the baryon number susceptibility~\cite{Borsanyi_order,Borsanyi_sus,Steinheimer}.  
It is also reported that, below the transition temperature,  the decrease of the absolute value of chiral condensate is well described by HRG+chiral perturbation theory ($\chi$PT)~\cite{Borsanyi_Tc}. 
Recently it was shown that the HRG model 
can also reproduce LQCD data on temperature dependence of 
the Polyakov-loop itself~\cite{Megias_PRL}.  
These results indicate that the effects of hadrons 
may be important in QCD phase transition, 
although these are not included in the simple effective model which treats 
the quark degrees of freedom only. 

In this paper, we  define the hadron-quark (confinement-deconfinement) transition temperature in the view of the ratio of quark and hadron contributions 
by using the simple hybrid model that has the mixture of quark and hadron matters, and separates the EOS into the quark and hadron components to see 
hadron effects on the transition. 
Our simple model can reproduce the 2+1 flavor lattice simulation data successfully.  
We also define the transition temperature of the transition by using the ratio of two phases. 
The temperature obtained from LQCD data on the EOS 
is about 215 MeV and is somewhat higher than 
the transition temperatures of chiral and $Z_3$ crossovers; 
here note that for each of chiral and $Z_3$ crossovers 
the transition temperature is usually  defined by 
the temperature at which 
the susceptibility or the absolute value of the derivative of the order parameter with respect to temperature becomes maximum.   

This paper as organized as follows. 
In Sec.~\ref{model}, we review the hadron resonance gas (HRG) model and also explain our quark-phase model with the Polyakov-loop. 
Then, we formulate the quark-hadron hybrid model which is used in this paper. 
Numerical results are shown in Sec~\ref{results}. 
Section~\ref{summary} is devoted to summary .
 
\section{Model}
\label{model}
\subsection{Hadron resonance gas model}

For pure hadronic matter, we use the HRG model. 
In the HRG model, the thermodynamic potential density of the system is given by the sum of free gas of hadron resonances. 
It is divided into two parts, namely, the baryonic and the mesonic parts.  
\begin{eqnarray}
{\rm \Omega_{\rm H}}={\rm \Omega_{\rm B}}+{\rm \Omega_{\rm M}}. 
\label{Omega_HRG}
\end{eqnarray}
The baryonic part ${\rm \Omega}_{\rm B}$ and the mesonic part ${\rm \Omega_{\rm M}}$ are given by
\begin{eqnarray}
{\rm \Omega_{\rm{B}}} &=& -\sum_{i \in \rm Baryon}d_{{\rm B},i}T\int \frac{d^3{\bf p}}{(2\pi)^3} \big\{ \log(1+ e^{-(E_{{\rm B},i} - \mu_{{\rm B},i})/T})\nonumber\\
&&+\log(1+ e^{-(E_{{\rm B},i} + \mu_{{\rm B},i})/T})\big\}; 
\nonumber\\
E_{{\rm B},i}&=&\sqrt{{\rm p}^2+{m_{{\rm B},i}}^2}, 
\label{Omega_B}
\end{eqnarray}
and 
\begin{eqnarray}
{\rm \Omega_{\rm{M}}} &=& \sum_{j \in \rm Meson}d_{{\rm M},j}T\int \frac{d^3{\bf p}}{(2\pi)^3} \big\{ \log(1- e^{-(E_{\rm M,j}-\mu_{{\rm M},j})/T})\nonumber\\
&&+\log(1- e^{-(E_{{\rm M},j}+\mu_{{\rm M},j})/T})\big\};
\nonumber\\
E_{{\rm M},j}&=&\sqrt{{\rm p}^2+{m_{{\rm M},j}}^2}, 
\label{Omega_M}
\end{eqnarray}
where $m_{{\rm B},i}$ ($m_{{\rm M},j}$) and $\mu_{{\rm B},i}$ ($\mu_{{\rm M},j}$) is the mass and the chemical potential of the $i$-th baryon ($j$-th meson) , respectively. 
The sums in Eq. (\ref{Omega_B}) and (\ref{Omega_M}) include all known baryons and mesons up to 2.5 GeV mass composed of u, d, s quarks, as listed in the latest edition of the Particle Data Book~\cite{PDG}. 

From the thermodynamic potential density, we obtain the pressure $P$ and the entropy density $s$ as follows. 
\begin{eqnarray}
P_{\rm H}&=&P_{\rm B}+P_{\rm M};~~~~~P_{\rm B}=-{\rm \Omega_{\rm B}},~~~~~P_{\rm M}=-{\rm \Omega_{\rm M}}, 
\label{P_H}
\\
s_{\rm H}&=&s_{\rm B}+s_{\rm M};~~~~~s_{\rm B}={\partial P_{\rm B}\over{\partial T}},~~~~~s_{\rm M}={\partial P_{\rm M}\over{\partial T}}. 
\label{s_H}
\end{eqnarray}

As mentioned in \S1, the HRG model can reproduce LQCD data well at low-temperature. 
In Ref.~\cite{Gorda}, the QCD equation of state was constructed by matching the low-temperature results of the HRG model with high temperature results of perturbative QCD (pQCD) at some intermediate temperature.  
Since we are interested in the relation between $T$ dependence of the Polyakov-loop and that of the deconfinement transition, for quark phase we do not use pQCD but an alternative model, as is described in the next subsection.

\subsection{Quark matter}

In this paper, as a model of pure quark matter, we use the quark model in which the current quark interacts with gluon field by gauge coupling. 
We neglect the spatial parts of gluon field and treat its temporal part $A_0$ as a stationary and uniform background field. 
We also use the gauge fixing in which $A_0$ is diagonal.  
Furthermore, instead of pure gluonic action, we use the effective potential of the Polyakov-loop. 
The Lagrangian density of this model is given by  
\begin{eqnarray}
{\cal L}_{\rm Q}=\sum_f \left\{\bar{q}_f(i\gamma^{\mu}D_{\mu}-m_f )q_f \right\}-{\cal U}(T,{\rm \Phi},{\rm \bar{\Phi}}), 
\label{L_Q}
\end{eqnarray}
where $D_{\mu}=\partial_{\mu}-igA_{\mu}^a\frac{\lambda_a}{2}$ with the Gell-Mann matrix $\lambda_a$ is the covariant derivative and reduces to
\begin{eqnarray}
\partial_{\mu}-ig\left( A_{0}^3\frac{\lambda_3}{2}+
A_{0}^8\frac{\lambda_8}{2}\right)\delta_{\mu, 0}
\label{covariant_PNJL}
\end{eqnarray}
 in our approximation mentioned above. We also emphasize that we do not include any condensate term in the Lagrangian~{(\ref{L_Q})}. As is shown in the next section, in the hadron resonances gas model, the chiral condensate is decreased rapidly by the hadron effects only and vanishes at $T~\sim~170$MeV below which the hadron phase dominate the EOS.  
Therefore, we do not include the chiral condensate term in quark matter. 
We do not include the other condensates, since the other condensates vanish at $\mu_{\rm B}=\mu_{\rm I}=0$.

The Polyakov-loop potential is given by 
\begin{eqnarray}
&&\frac{{\cal U}(T,{\rm \Phi},{\rm \bar{\Phi}})}{T^4}=-\frac{a(T)}{2}{\rm \Phi}{\rm \bar{\Phi}}\nonumber\\
&&+ b(T)\log\{1-6{\rm \Phi}{\rm \bar{\Phi}} + 4({\rm \Phi}^3 + {\rm \bar{\Phi}}^3) - 3({\rm \Phi} {\rm \bar{\Phi}})^2\};
\label{U_Phi}\\
&&a(T) = a_0 + a_1\left(\frac{T_0}{T}\right)+ a_2\left(\frac{T_0}{T}\right)^2,\\
&&b(T) = b_3\left(\frac{T_0}{T}\right)^3, 
\label{parameter_Phi}
\end{eqnarray}
where $a_0$, $a_1$, $a_2$, $b_3$ and $T_0$ are the constant parameters. 
The PNJL model is expected to reduce to this model at high temperature region where the quark-quark direct interactions vanishes.   
Hence, we use the same parameter set $a_0$, $a_1$, $a_2$, $b_3$ and $T_0$ as in the PNJL model. 
The values of these parameters~\cite{Rossner} are summarized in Talbe~\ref{Table_cal_U}.

\begin{table}[h]
\centering
\begin{tabular}{lcccr}
\hline
$\ a_0$&$a_1$&$a_2$&$b_3$&$T_0\hspace{5mm}$
\\ \hline
3.51&-2.47&15.2&-1.75&270{\rm [MeV]}\\
\hline
\end{tabular}
\caption{Parameters of Polyakov-loop potential. }
\label{Table_cal_U}
\end{table}

\begin{figure}[h]
\begin{center}
\vspace{0.5cm}
\includegraphics[width=0.35\textwidth]{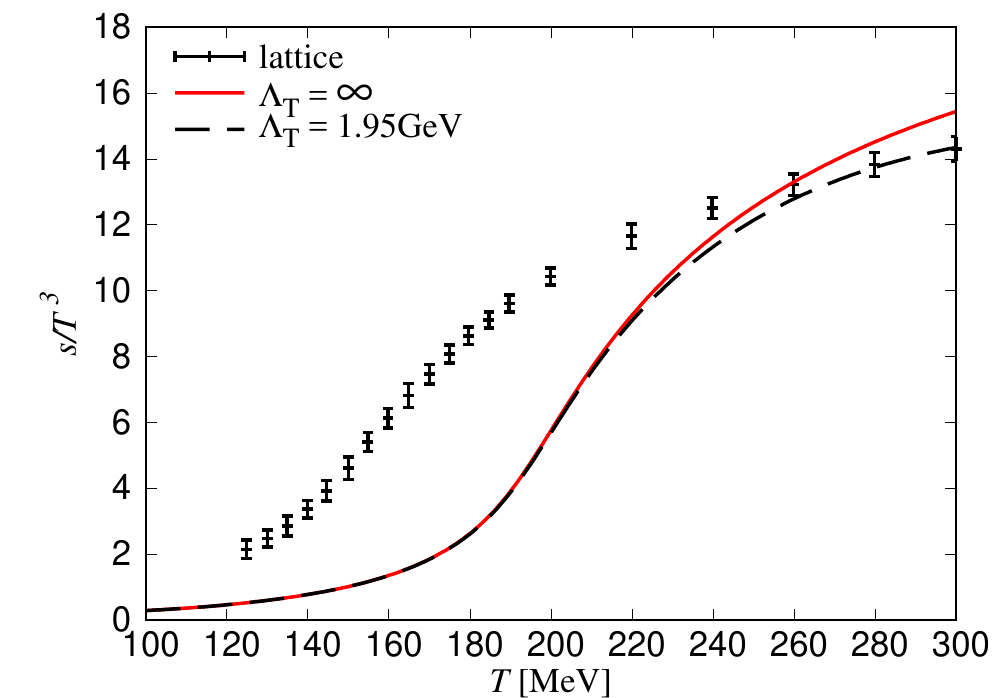}
\end{center}
\vspace{10pt}
\caption{$T$-dependence of the entropy density at $\mu_{\rm B}=\mu_{\rm I}=0$. 
The results were obtained by using the quark-phase model with $\Lambda_{\rm T}=$1.95GeV and $\infty$, respectively. 
LQCD date are taken from Ref.~\cite{Borsanyi_entropy}.
}
\label{entropy_add}
\end{figure}

Using the Lagrangian (\ref{L_Q}), we obtain the thermodynamic potential density of the quark phase. 
\begin{eqnarray}
{\rm \Omega}_{\rm Q}&=&\ {\cal U}(T,{\rm \Phi},{\rm \bar{\Phi}}) 
- 2\sum_{f=u,d,s} \int_{|{\bf p}|\leq \Lambda} \frac{d^3{\bf p}}{(2\pi)^3} \ 3E_f\nonumber\\
&&-2\sum_{f=u,d,s} \int_{|{\bf p}|\leq \Lambda_{\rm T}} \frac{d^3{\bf p}}{(2\pi)^3} ( T\log{z_f^+}+ T\log{z_f^-}), 
\nonumber\\
\label{Omega_Q}
\end{eqnarray}
where 
\begin{eqnarray}
z_f^+ &=& 1+3{\rm \bar{\Phi}} e^{-(E_f+\mu_f)/T}+3{\rm \Phi} e^{-2(E_f+\mu_f)/T} 
\nonumber\\
&&+ e^{-3(E_f+\mu_f)/T},
\label{zfp}\\
z_f^- &=& 1+3{\rm \Phi} e^{-(E_f-\mu_f)/T}+3{\rm \bar{\Phi}} e^{-2(E_f-\mu_f)/T}
\nonumber\\
&&+ e^{-3(E_f-\mu_f)/T};
\label{zfm}\\
&&E_f=\sqrt{{\bf p}^2+m_f^2} ,
\label{eff}
\end{eqnarray}
with the current quark mass $m_f$ of the $f$ flavor quark. 
In Eq.~(\ref{Omega_Q}), $\Lambda_{\rm T}$ is the phenomenological cutoff of thermal excitation term in ${\rm \Omega_Q}$ and determined to reproduce  the entropy density at $T=300$MeV obtained by LQCD simulations. 
The obtained value is $\Lambda_{\rm T} = 1.95{\rm GeV}$. 
In Fig.~{\ref{entropy_add}}, we show the entropy densities calculated by using our quark-phase models with $\Lambda_{\rm T}=1.95$GeV and $\Lambda_{\rm T}=\infty$, respectively. 
The model with $\Lambda_{\rm T}=1.95$GeV reproduces LQCD data~\cite{Borsanyi_entropy} well at high temperature, while the model with $\Lambda_{\rm T}=\infty$ overshoots the LQCD data. 
Therefore, the cutoff is necessary in our model.
We emphasize that $\Lambda_{\rm T}$ is another parameter that is being used as a cutoff to match with LQCD data.
If the ideal free gas is considered,  the thermal distribution function 
will give 
a natural cutoff in the thermal contribution of the thermodynamical potential density, and therefore $\Lambda_{\rm T}$ will go infinity in that limit. 
Our quark-phase model is not an ideal-gas model but an 
effective model with the phenomenological parameter $\Lambda_{\rm T}$ that 
is introduced to fit LQCD data.
There is also a cutoff $\Lambda$ in the second term (vacuum contribution term) in the right-hand-side of Eq.~(\ref{Omega_Q}). 
However, in our model, this term becomes constant and consequently  is not relevant in our analyses on only temperature and chemical potential dependences of the physical quantities. 
Below, we omit this vacuum term in our analyses; hence we do not  determine $\Lambda$.

From the thermodynamic potential density, we obtain the pressure $P_{\rm Q}$ and the entropy density $s_{\rm Q}$ as follows. 
\begin{eqnarray}
P_{\rm Q}&=&-{\rm \Omega_{\rm Q}}, 
\label{P_Q}
\\
s_{\rm Q}&=&{\partial P_{\rm Q}\over{\partial T}}. 
\label{s_Q}
\end{eqnarray}

In this subsection, we use the flavor-dependent quark number chemical potential $\mu_f~(f=u,d,s)$~\cite{Allton_chem,Andersen_chem} just for simplicity. 
In literatures of LQCD calculations,  
the baryonic chemical potential $\mu_{\rm B}$, the isospin chemical potential $\mu_{\rm I}$ and the strangeness chemical potential $\mu_{\rm S}$ are often used. 
The relation of these quantities is 
\begin{eqnarray}
\mu_u={1\over{3}}\mu_{\rm B}+\frac{1}{2}\mu_{\rm I}, ~~~\mu_d={1\over{3}}\mu_{\rm B}-\frac{1}{2}\mu_{\rm I},~~~\mu_s={1\over{3}}\mu_{\rm B}-\mu_{\rm S}.\nonumber\\
\label{chemical}
\end{eqnarray}
Below, we use $\mu_{\rm B}$ and $\mu_{\rm I}$ instead of $\mu_f$ 
by setting $\mu_{\rm S}=0$.  

\subsection{Quark-hadron hybrid model}

In our calculation, we use the quark-hadron hybrid model~\cite{Hatsuda_hybrid model} which is composed of quark and hadronic matter. 
In the model, the total entropy density of the system is given by  
\begin{eqnarray}
s(T,\mu_{\rm B}^2,\mu_{\rm I}^2) &=& (1-f_{\rm H}(T,\mu_{\rm B}^2,\mu_{\rm I}^2))s_{\rm Q}(T,\mu_{\rm B}^2,\mu_{\rm I}^2)\nonumber\\
&&+ f_{\rm H}(T,\mu_{\rm B}^2,\mu_{\rm I}^2)s_{\rm H}(T,\mu_{\rm B}^2,\mu_{\rm I}^2), 
\label{s_Hybrid}
\end{eqnarray}
where $f_{\rm H}(T,\mu_{\rm B}^2,\mu_{\rm I}^2)$ is the hadron volume fraction function. 
The system is pure hadronic matter (quark matter) when $f_{\rm H}(T,\mu_{\rm B}^2,\mu_{\rm I}^2)=1~(0)$.  
Using the total entropy density $s$, pressure is obtained as 
\begin{eqnarray}
&&P(T,\mu_{\rm B}^2,\mu_{\rm I}^2)-P(0,\mu_{\rm B}^2,\mu_{\rm I}^2)\nonumber\\
&=&\int_{0}^{T}dT' s(T',\mu_{\rm B}^2,\mu_{\rm I}^2)\nonumber\\
&=&[P_{{\rm Q}}]_{0}^T - \int_{0}^{T}dT' f_{\rm H}(T',\mu_{\rm B}^2,\mu_{\rm I}^2)s_{\rm Q}(T',\mu_{\rm B}^2,\mu_{\rm I}^2)\nonumber\\
&&+\int_{0}^{T}dT'f_{\rm H}(T',\mu_{\rm B}^2,\mu_{\rm I}^2)s_{\rm H}(T',\mu_{\rm B}^2,\mu_{\rm I}^2).
\label{P_Hybrid}
\end{eqnarray}
Using $P$, the baryon number susceptibility is defined as the second derivative of $P$ with respect to the baryonic chemical potential $\mu_{\rm B}$. 
\begin{eqnarray}
&&\chi_{\rm B}(T,\mu_{\rm B}^2,\mu_{\rm I}^2)-\chi_{\rm B}(0,\mu_{\rm B}^2,\mu_{\rm I}^2)\nonumber\\
&=&\frac{\partial^2}{\partial \mu_{\rm B}^2} (P(T,\mu_{\rm B}^2,\mu_{\rm I}^2)-P(0,\mu_{\rm B}^2,\mu_{\rm I}^2))\nonumber\\
&=& [\chi_{\rm B}^{{\rm Q}}]_{0}^T + \int_0^T dT' \{\frac{\partial^2 f_{\rm H}}{\partial \mu_{\rm B}^2}(s_{\rm H}-s_{\rm Q})\nonumber\\
&&+2\frac{\partial f_{\rm H}}{\partial \mu_{\rm B}}\frac{\partial (s_{\rm H}-s_{\rm Q})}{\partial \mu_{\rm B}} + f_{\rm H}\frac{\partial^2 (s_{\rm H}-s_{\rm Q})}{\partial \mu_{\rm B}^2}\}\label{chi_B}. 
\end{eqnarray}
In particular at $\mu_{\rm B}=0$, one obtain 
\begin{eqnarray}
&&\chi_{\rm B}(T)-\chi_{\rm B}(0)\nonumber\\
&=& [\chi_{\rm B}^{{\rm Q}}]_{0}^T + \int_0^T dT' \{2\frac{\partial f_{\rm H}}{\partial \nu_{\rm B}}(s_{\rm H}-s_{\rm Q})+ f_{\rm H}\frac{\partial^2 (s_{\rm H}-s_{\rm Q})}{\partial \mu_{\rm B}^2}\}\nonumber\\
&=& [\chi_{\rm B}^{{\rm Q}}]_{0}^T - \int_0^T dT' \Bigg\{2\frac{\partial f_{\rm H}}{\partial \nu_{\rm B}}s_{\rm Q}+ f_{\rm H}\frac{\partial^2 s_{\rm Q}}{\partial \mu_{\rm B}^2} \Bigg\}\nonumber\\
&&+\int_0^TdT^\prime \Bigg\{2\frac{\partial f_{\rm H}}{\partial \nu_{\rm B}}s_{\rm H}+ f_{\rm H}\frac{\partial^2 s_{\rm H}}{\partial \mu_{\rm B}^2} \Bigg\}, 
\label{chi_B_mu=0}
\end{eqnarray}
where $\nu_{\rm B} \equiv \mu_{\rm B}^2$.
Similarly, the isospin number susceptibility is given by the following equation. 
\begin{eqnarray}
&&\chi_{\rm I}(T,\mu_{\rm B}^2,\mu_{\rm I}^2)-\chi_{\rm I}(0,\mu_{\rm B}^2,\mu_{\rm I}^2)\nonumber\\
&=&\frac{\partial^2}{\partial \mu_{\rm I}^2} (P(T,\mu_{\rm B}^2,\mu_{\rm I}^2)-P(0,\mu_{\rm B}^2,\mu_{\rm I}^2))\nonumber\\
&=& [\chi_{\rm I}^{{\rm Q}}]_{0}^T + \int_0^T dT' \{\frac{\partial^2 f_{\rm H}}{\partial \mu_{\rm I}^2}(s_{\rm H}-s_{\rm Q})\nonumber\\
&&+2\frac{\partial f_{\rm H}}{\partial \mu_{\rm I}}\frac{\partial (s_{\rm H}-s_{\rm Q})}{\partial \mu_{\rm I}} + f_{\rm H}\frac{\partial^2 (s_{\rm H}-s_{\rm Q})}{\partial \mu_{\rm I}^2}\} .
\label{chi_iso}
\end{eqnarray}
At $\mu_{\rm I}=0$, 
\begin{eqnarray}
&&\chi_{\rm I}(T)-\chi_{\rm I}(0)\nonumber\\
&=& [\chi_{\rm I}^{{\rm Q}}]_{0}^T - \int_0^T dT' \Bigg\{2\frac{\partial f_{\rm H}}{\partial \nu_{\rm I}}s_{\rm Q}+ f_{\rm H}\frac{\partial^2 s_{\rm Q}}{\partial \mu_{\rm I}^2} \Bigg\}\nonumber\\
&&+\int_0^TdT^\prime \Bigg\{2\frac{\partial f_{\rm H}}{\partial \nu_{\rm I}}s_{\rm H}+ f_{\rm H}\frac{\partial^2 s_{\rm H}}{\partial \mu_{\rm I}^2} \Bigg\}, 
\label{chi_iso_mui=0}
\end{eqnarray}
where $\nu_{\rm I}={\mu_{\rm I}}^2$. 

We also calculate the $T$-dependence of the chiral condensate $\sigma_f$ of $f$ flavor quark by using the following equation. 
\begin{eqnarray}
&&\sigma_{f}(T,\mu_{\rm B}^2,\mu_{\rm I}^2)-\sigma_{f}(0,\mu_{\rm B}^2,\mu_{\rm I}^2)\nonumber\\
&=&\frac{\partial {\rm \Omega}}{\partial m_{f}}(T,\mu_{\rm B}^2,\mu_{\rm I}^2)-\frac{\partial {\rm \Omega}}{\partial m_{f}}(0,\mu_{\rm B}^2,\mu_{\rm I}^2)\nonumber\\
&=& -\frac{\partial}{\partial m_{f}}\int_0^{T}dT'\left[(1-f_{\rm H})s_{\rm Q}+f_{\rm H}s_{\rm H}\right]\nonumber\\
&=& [\sigma^{{\rm Q}}_{f}]_{0}^T + \int_0^{T}dT'\left[ f_{\rm H}\left(\frac{\partial \sigma^{\rm H}_{f}}{\partial T'}-\frac{\partial \sigma^{\rm Q}_{f}}{\partial T'}\right)\right],
\label{sigma_f_T}
\end{eqnarray}
where we use $\frac{\partial s}{\partial m} = -\frac{\partial}{\partial m}\frac{\partial {\rm \Omega}}{\partial T} = -\frac{\partial}{\partial T}\frac{\partial {\rm \Omega}}{\partial m} = -\frac{\partial \sigma}{\partial T}$. 
In Eq.~(\ref{sigma_f_T}),  
\begin{eqnarray}
\sigma^{\rm H}_f &=& \frac{\partial {\rm \Omega_{\rm H}}}{\partial m_f}\nonumber\\
&=& \sum_{i \in \rm Baryon} \frac{\partial M_{{\rm B},i}}{\partial m_f}\frac{\partial {\rm \Omega_{\rm H}}}{\partial M_{{\rm B},i}} + \sum_{j \in \rm Meson} \frac{\partial M_{{\rm M},j}}{\partial m_f}\frac{\partial {\rm \Omega_{\rm H}}}{\partial M_{{\rm M},j}}, 
\nonumber\\
\label{sigma_f_h}
\end{eqnarray}
where 
\begin{eqnarray}
\frac{\partial M_{{\rm B},i}}{\partial m_f} = C_f^{{\rm B},i}, 
\label{CgB}
\\
\frac{\partial M_{{\rm M},j}}{\partial m_f} = C_f^{{\rm M},j}. 
\label{CgM}
\end{eqnarray}
The coefficients $C_f^{{\rm B},i}$ and $C_f^{{\rm M},j}$ are the constants decided by the number of  $f$ quark included in the hadron $M_{{\rm B},i}$ and $M_{{\rm M},j}$ which are not the octet Nambu-Goldstone (NG) bosons. 
For example, in the case of proton composed of two u quarks and one d quark, $C_u^{\rm p} = 2$, $C_d^{\rm p}= 1$ and $C_s^{\rm p}= 0$. 
For the octet NG bosons, $\pi$, $K$ and $\eta$, the Gell-Mann-Oakes-Renner (GMOR) relation~\cite{GMOR relation} 
is used to determine $C_f^{{\rm M},j}$. 

Using $\sigma_{f}$ renormalized chiral condensate is given by
\begin{eqnarray}
\Delta_{l,s}(T,\mu_{\rm B}^2,\mu_{\rm I}^2) \equiv \frac{\sigma_l(T,\mu_{\rm B}^2,\mu_{\rm I}^2)-(\frac{m_l}{m_s})\sigma_s(T,\mu_{\rm B}^2,\mu_{\rm I}^2)}{\sigma_l(0,\mu_{\rm B}^2,\mu_{\rm I}^2)-(\frac{m_l}{m_s})\sigma_s(0,\mu_{\rm B}^2,\mu_{\rm I}^2)},\nonumber\\ 
\label{Delta_ls}
\end{eqnarray}
where $m_l = \frac{m_u+m_d}{2}$ is the average value of current quark mass of light quarks.  
Hereafter, we put $m_u=m_d=m_l$.  
Using a quantity 
\begin{eqnarray}
\Sigma_{f}(T) &\equiv& -\frac{\partial}{\partial m_f}\int_0^{T}dT'\left[(1-f_{\rm H})s_{\rm Q}+f_{\rm H}s_{\rm H}\right]
\nonumber\\
&=& \sigma_{f}(T)-\sigma_{f}(0),
\label{Large_Sigma}
\end{eqnarray}
Eq.~(\ref{Delta_ls}) is rewritten as 
\begin{eqnarray}
\Delta_{l,s}(T)=1+\frac{\Sigma_l(T)-(\frac{m_l}{m_s})\Sigma_s(T)}{\sigma_l(0)-(\frac{m_l}{m_s})\sigma_s(0)}, 
\label{Delta_ls_rw}
\end{eqnarray}
where $\sigma_l(0), \sigma_s(0)$ is derived by GMOR relation for $\pi$ and $K$.

\section{Numerical results}
\label{results}

In this section, we determine the hadron volume fraction function $f_{\rm H}$ to reproduce the 2+1 flavor LQCD data on the entropy density, the baryon number and the isospin number susceptibilities. 
Using the $f_{\rm H}$ thus determined, we examine how our hybrid model reproduces the other thermodynamical quantities of LQCD. 
We use LQCD data of Refs.~\cite{Borsanyi_entropy, Borsanyi_sus_plot, Borsanyi_order}, since the data are available for all thermodynamical quantities needed for our analyses and a lot of numerical data are published in addition to the graphical ones. 
Although more recent data~\cite{Borsanyi_sus} were presented by the same group, we found the data are too limited to perform the present analysis. 
Hence we did not use the latest data in the present analyses. 
There are also the 2+1+1 LQCD data~\cite{Bellwied_2+1+1} by the same group. 
Analyses of these new data are interesting future works.

{\subsection{Entropy density and hadron volume fraction function}

In Fig.~\ref{entropy}, we show the entropy density $s$ of the quark-hadron hybrid model.  
The hadron volume fraction function $f_{\rm H}$ at $\mu_{\rm B}=\mu_{\rm I}=0$ is determined to reproduce the LQCD data of $s$~\cite{Borsanyi_entropy}.  
Here, the temperature dependence of  it is shown in Fig.~\ref{fH0}. 
The explicit form of $f_{\rm H}$ is given by following equation. 
\begin{eqnarray}
f_{\rm H}(T,0,0) = \frac{1}{2}\left\{1+\tanh{((b - T)e^{\left(\frac{c}{T}\right)^d}/a)}\right\}, 
\label{function of fH0}
\end{eqnarray}
where the parameters are tabulated in Table~\ref{parameters of fH0}. 
One may try to fit $f_{\rm H}(T,0,0)$ by a simpler function 
\begin{eqnarray}
f_{\rm H}(T,0,0) = \frac{1}{2}\left\{1+\tanh{((b - T)/a}\right\}.  
\label{simpler_fH}
\end{eqnarray}
However, Eq. (\ref{simpler_fH}) gives the nontrivial antisymmetric relation $f_{\rm H}(b-\Delta T,0,0)-1/2=-(f_{\rm H}(b+\Delta T,0,0)-1/2)$ that LQCD data dose not have correctly.  
Hence, we add the additional factor $e^{\left(\frac{c}{T}\right)^d}$ that breaks the antisymmetric relation. 
 
\begin{table}[h]
\centering
\begin{tabular}{lccr}
\hline
$\hspace{10mm}a$&$\hspace{5mm}b$&$c$&$d\hspace{5mm}$
\\ \hline
27.0326{\rm [MeV]}&205.458{\rm [MeV]}&174.154{\rm [MeV]}&17\hspace{5mm}
\\ \hline
\end{tabular}
\caption{Parameters of $f_{\rm H}$.}
\label{parameters of fH0}
\end{table}

In Fig.~\ref{entropy}, 
the entropy density is divided into hadron and quark parts. 
The critical temperature $T_c^{(s)}$ of the quark-hadron transition is defined as the temperature at the crosspoint of the quark and hadron parts.  
The value is $T_c^{(s)}=215$MeV. 

\begin{figure}[h]
\begin{center}
\vspace{0.5cm}
\includegraphics[width=0.35\textwidth]{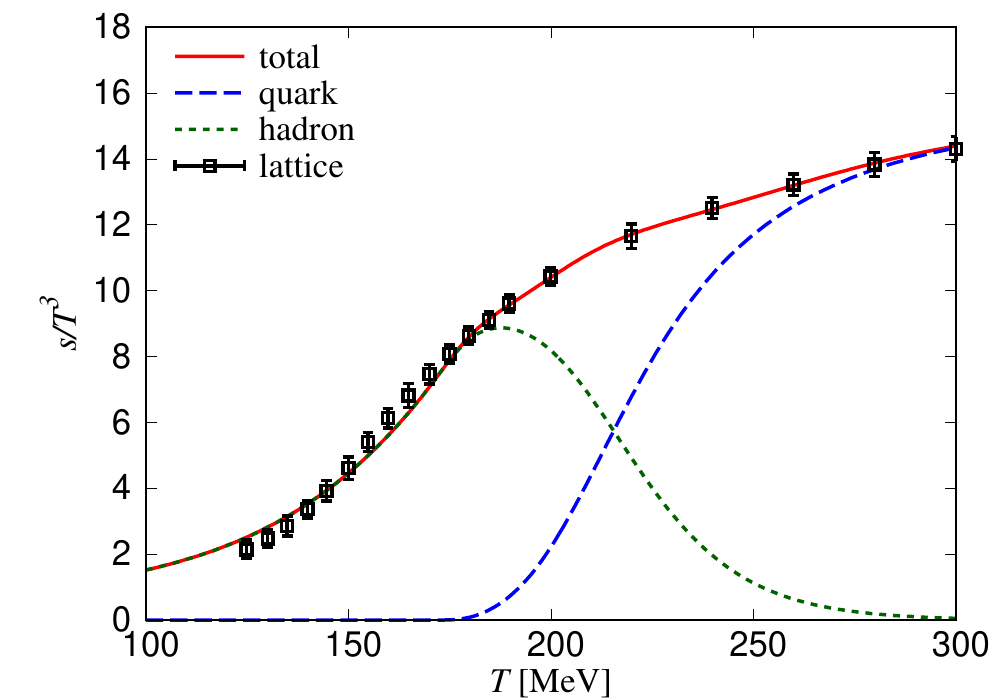}
\end{center}
\vspace{10pt}
\caption{$T$-dependence of the entropy density at $\mu_{\rm B}=\mu_{\rm I}=0$. 
The LQCD date is taken from Ref.~\cite{Borsanyi_entropy}. 
We show $s_{\rm H}f_{\rm H}$ ($s_{\rm Q}(1-f_{\rm H})$) as a hadron (quark) contribution. 
These quantities do not diverge at high temperature.
}
\label{entropy}
\end{figure}
\begin{figure}[h]
\begin{center}
\vspace{0.5cm}
\includegraphics[width=0.35\textwidth]{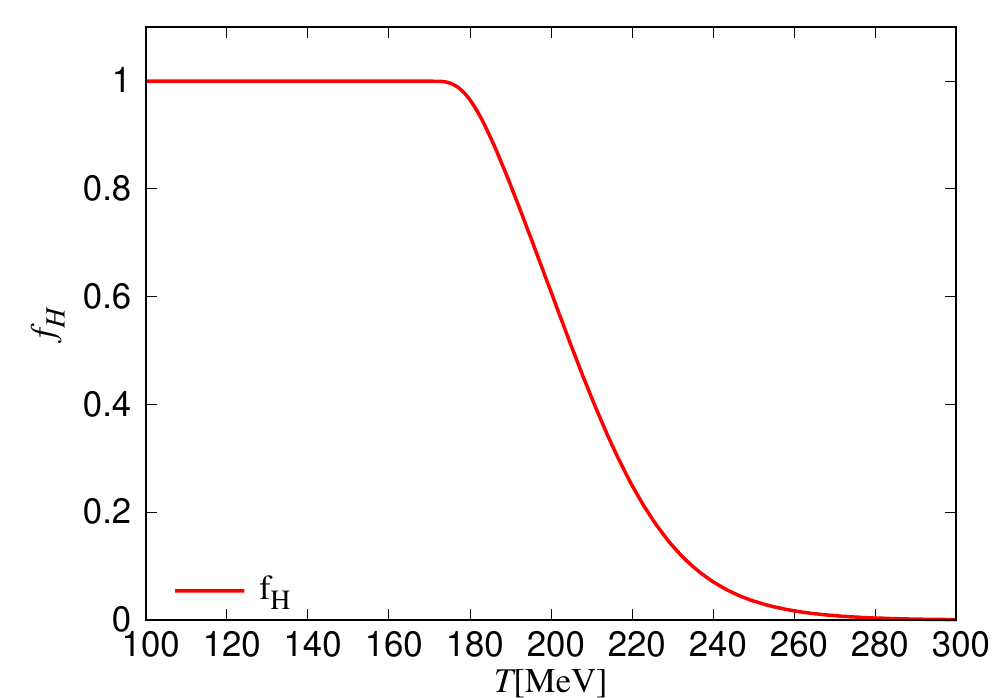}
\end{center}
\vspace{10pt}
\caption{$T$-dependence of the $f_H$ at $\mu_{\rm B}=\mu_{\rm I}=0$. }
\label{fH0}
\end{figure}

\subsection{Pressure}

Figure~\ref{pressure} shows $T$-dependence of the total pressure $P$ at $\mu_{\rm B}=\mu_{\rm I}=0$. 
The hybrid model reproduces the pressure obtained by LQCD simulation~\cite{Borsanyi_entropy} and is possible to be divided into the contributions of hadron and quark parts. 
In the context of two phase model of quark-hadron phase transition~\cite{Cleymans}, 
the transition temperature is defined as the temperature where the pressure of quark phase is  equal to that of the hadron phase.  
When we adopt the same definition of transition temperature, namely, $P_{\rm Q}=P_{\rm H}$, in our model, 
we obtained $T_c^{(P)}=249$MeV which is somewhat larger than $T_c^{(s)}$. 

\begin{figure}[h]
\begin{center}
\vspace{0.5cm}
\includegraphics[width=0.35\textwidth]{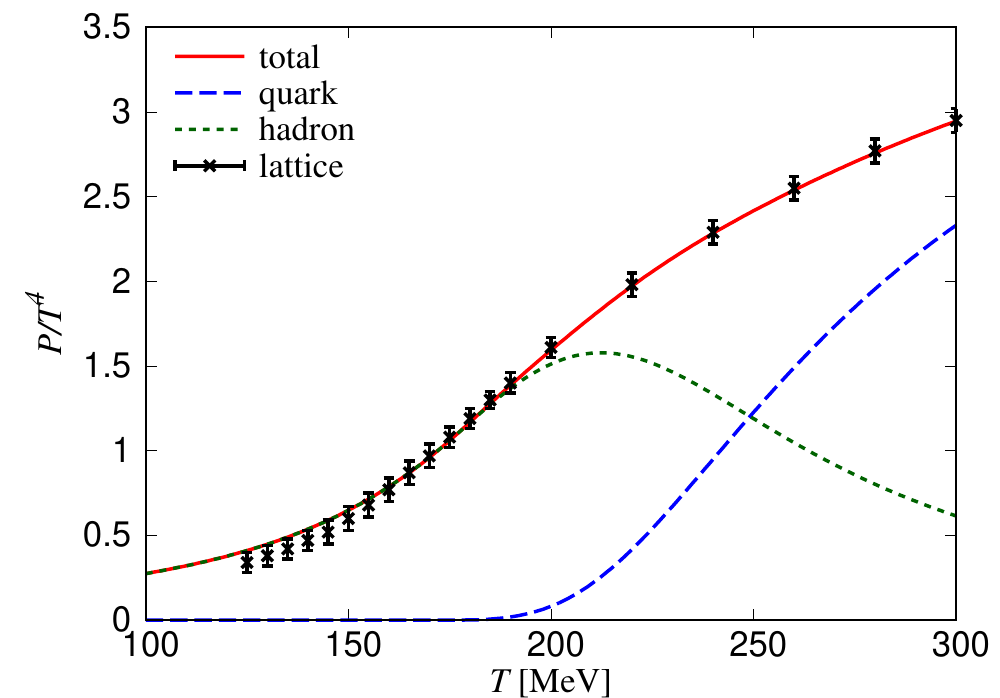}
\end{center}
\vspace{10pt}
\caption{$T$-dependence of the pressure at $\mu_{\rm B}=\mu_{\rm I}=0$. 
We put $P=0$ at $T=\mu_{\rm B}=\mu_{\rm I}=0$. 
The LQCD date is taken from Ref.~\cite{Borsanyi_entropy}. 
We show the third (first + second) term in the right-hand-side of Eq.~(\ref{P_Hybrid}), as a hadron (quark) contribution. 
These quantities do not diverge at high temperature.}
\label{pressure}
\end{figure}

\subsection{Interaction measure} 

In Fig. \ref{trace anomaly}, we show interaction measure (trace anomaly) at $\mu_{\rm B}=\mu_{\rm I}=0$. {Again, our simple hybrid model can reproduce the LQCD data very well~\cite{Borsanyi_entropy}.  
The interaction measure has a maximum around $T=T_{\rm max}^{\rm Int,L}=200$MeV. 
 It is interesting that $T_c^{(s)}$ is closer to $T_{\rm max}^{\rm Int,L}$ rather than $T_c^{C,{\rm L}}$ and $T_c^{Z_3,{\rm L}}$. } 
Furthermore, the temperature $T_{1/2}^{f_{\rm H}}$ where $f_H$ is equal to 1/2 coincides almost with $T_{ \rm max}^{\rm Int,L}$. 

\begin{figure}[h]
\begin{center}
\vspace{0.5cm}
\includegraphics[width=0.35\textwidth]{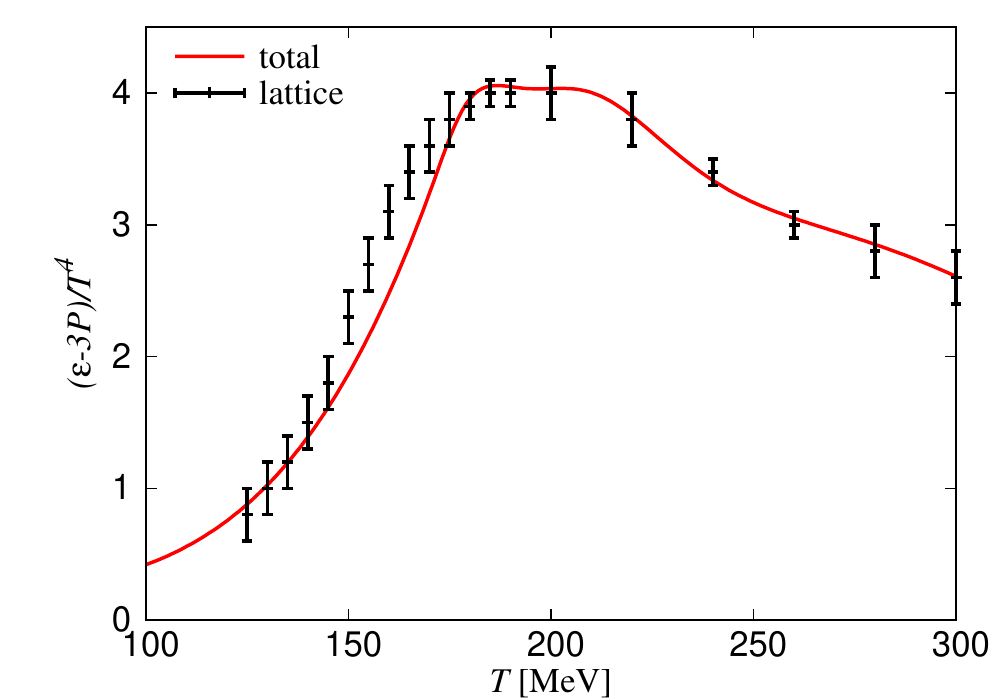}
\end{center}
\vspace{10pt}
\caption{$T$-dependence of the interaction measure at $\mu_{\rm B}=\mu_{\rm I}=0$. 
The LQCD date is taken from Ref.~\cite{Borsanyi_entropy}. }
\label{trace anomaly}
\end{figure}
\vspace{-5.3mm}
\subsection{Baryon and isospin number susceptibilities}

At $\mu_{\rm B}=\mu_{\rm I}=0$, we determine the derivatives $\frac{\partial f_{\rm H}}{\partial \nu_{\rm B}}$ and $\frac{\partial f_{\rm H}}{\partial \nu_{\rm I}}$ in Eqs.~(\ref{chi_B_mu=0}) and (\ref{chi_iso_mui=0}) to reproduce the baryon and isospin susceptibilities in LQCD simulations~\cite{Borsanyi_sus_plot}. 
We assume the explicit form of $\frac{\partial f_{\rm H}}{\partial \nu_{\rm B}}$ and $\frac{\partial f_{\rm H}}{\partial \nu_{\rm I}}$ as  
\begin{eqnarray}
\frac{\partial f_{\rm H}}{\partial \nu_B}(T,0,0) &=& a_{\rm B}\exp{\Bigg[- \left(\frac{T-b_{\rm B}}{c_{\rm B}}\right)^2\Bigg]}
\label{f_H_B}, 
\\
\frac{\partial f_{\rm H}}{\partial \nu_I}(T,0,0) &=& a_{\rm I}\exp{\Bigg[- \left(\frac{T-b_{\rm I}}{c_{\rm I}}\right)^2\Bigg]}, 
\label{f_H_I}
\end{eqnarray}
and search a parameter set which reproduces the LQCD simulation well~\cite{Borsanyi_sus_plot}. 
We found that the simple Gaussian forms of (\ref{f_H_B}) and (\ref{f_H_I}) are adequate to reproduce the data. 
The obtained parameters are tabulated in Table~\ref{parameters of fH2B} and Table~\ref{parameters of fH2I}.  
The obtained $\frac{\partial f_{\rm H}}{\partial \nu_{\rm B}}$ and $\frac{\partial f_{\rm H}}{\partial \nu_{\rm I}}$ are shown in Fig.~\ref{fH2}, and the reproduced  $\chi_{\rm B}$ and $\chi_{\rm I}$ are shown in Figs.~\ref{baryon number susceptibility} and \ref{isospin number susceptibility}, respectively. 

\begin{table}[h]
\centering
\begin{tabular}{lcr}
\hline
$\hspace{10mm}a_{\rm B}$&$\hspace{5mm}b_{\rm B}$&$c_{\rm B}\hspace{5mm}$
\\ \hline
-3.74513${\rm [MeV]^{-2}}$&192{\rm [MeV]}&30.6622{\rm [MeV]}
\\ \hline
\end{tabular}
\caption{Parameters of $\frac{\partial f_{\rm H}}{\partial \nu_{\rm B}}$.}
\label{parameters of fH2B}
\end{table}

\begin{table}[h]
\centering
\begin{tabular}{lcr}
\hline
$\hspace{10mm}a_{\rm I}$&$\hspace{5mm}b_{\rm I}$&$c_{\rm I}\hspace{5mm}$
\\ \hline
-5.2969${\rm [MeV]^{-2}}$&181.929{\rm [MeV]}&36.3811{\rm [MeV]}
\\ \hline
\end{tabular}
\caption{Parameters of $\frac{\partial f_{\rm H}}{\partial \nu_{\rm I}}$.}
\label{parameters of fH2I}
\end{table}

\begin{figure}[h]
\begin{center}
\vspace{0.5cm}
\includegraphics[width=0.35\textwidth]{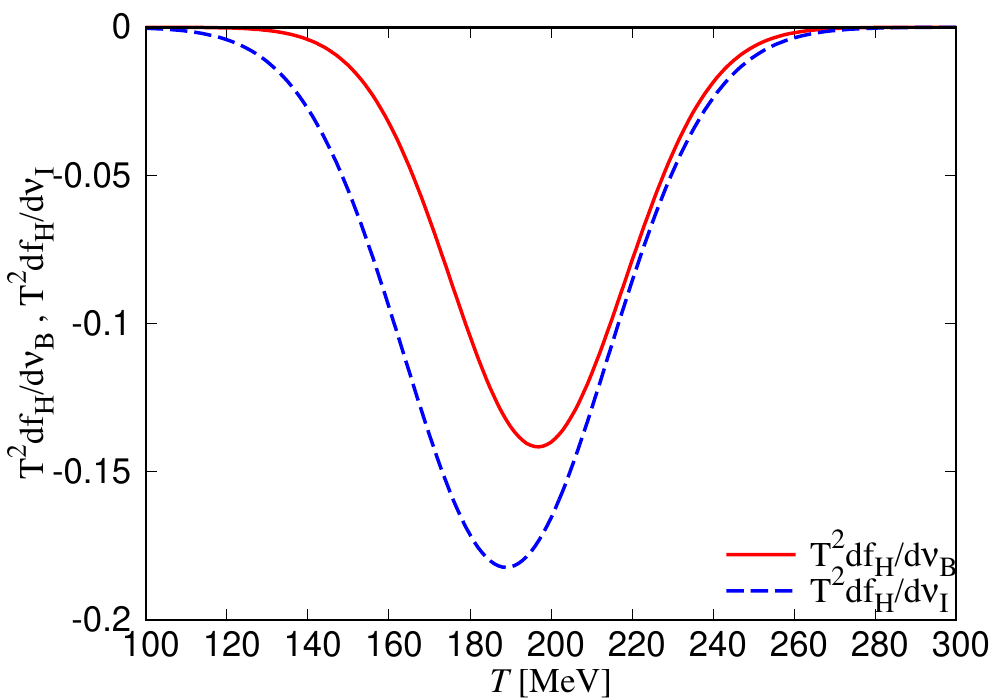}
\end{center}
\vspace{10pt}
\caption{$T$-dependence of $\frac{\partial f_H}{\partial \nu_B},\frac{\partial f_H}{\partial \nu_I}$ at $\mu_{\rm B}=\mu_{\rm I}=0$.}
\label{fH2}
\end{figure}

\begin{figure}[h]
\begin{center}
\vspace{0.5cm}
\includegraphics[width=0.35\textwidth]{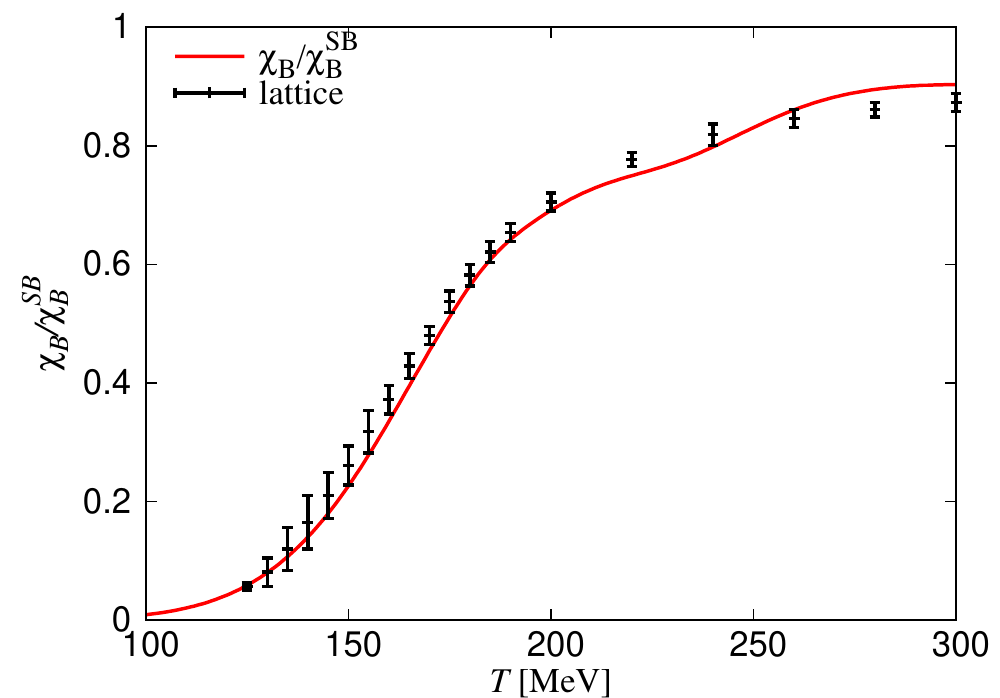}
\end{center}
\vspace{10pt}
\caption{$T$-dependence of the baryon number susceptibility at $\mu_{\rm B}=\mu_{\rm I}=0$. 
The $\chi_{\rm B}$ is normalized by its Stefan Boltzmann (SB) limit. 
The LQCD date is taken from Ref.~\cite{Borsanyi_sus_plot}. }
\label{baryon number susceptibility}
\end{figure}
\begin{figure}[h]
\begin{center}
\vspace{0.5cm}
\includegraphics[width=0.35\textwidth]{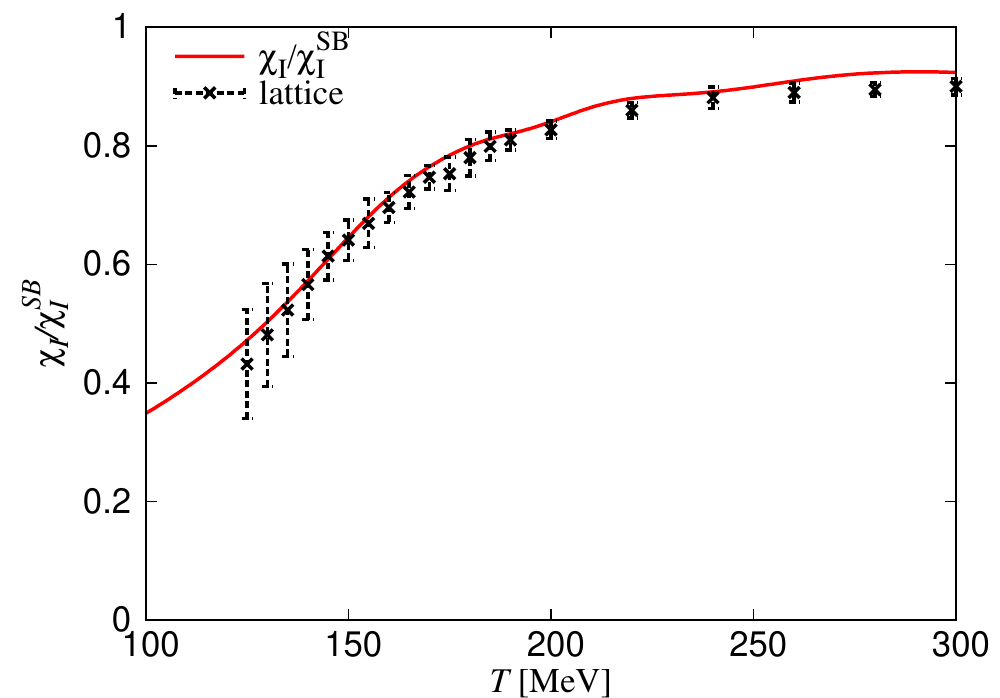}
\end{center}
\vspace{10pt}
\caption{$T$-dependence of the isospin number susceptibility at $\mu_{\rm B}=\mu_{\rm I}=0$. 
The $\chi_{\rm B}$ is normalized by its SB limit. 
The LQCD date is taken from Ref.~\cite{Borsanyi_sus_plot}. }
\label{isospin number susceptibility}
\end{figure}

\subsection{Polyakov-loop}

In Fig.~\ref{Fig_polyakov loop}, we show Polyakov-loop, the order parameter of $Z_3$ transition, at $\mu_{\rm B}=\mu_{\rm I}=0$.
Our model can reproduce the LQCD result~\cite{Borsanyi_order} very well up to $T=190$MeV but deviates from it at higher temperature. 
We define the transition temperature $T_c^{Z_3}$ of $Z_3$ transition as the temperature where the derivative ${d{\rm \Phi}\over{dT}}$ has its maximum. 
The obtained value $T_c^{Z_3}=198$MeV is somewhat larger than that $T_c^{Z_3,{\rm L}}=170\pm 7$MeV in LQCD calculation~\cite{YAoki_Tc^(Z_3)}, but is rather smaller than $T_{c}^{ (s)}$  and $T_{c}^{(P)}$. 
\begin{figure}[h]
\begin{center}
\vspace{0.5cm}
\includegraphics[width=0.35\textwidth]{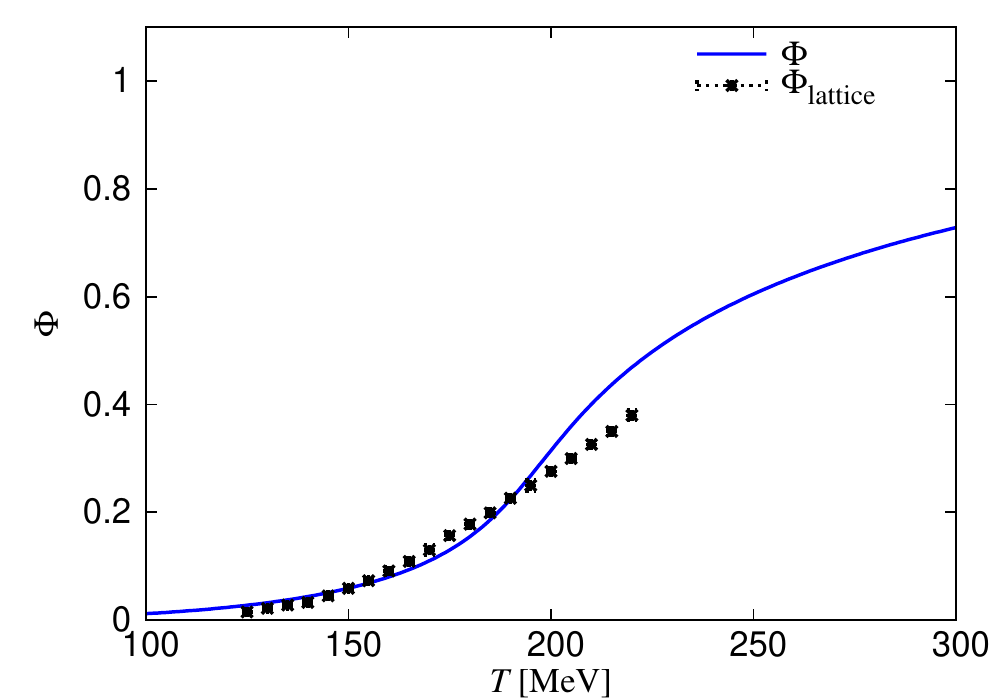}
\end{center}
\vspace{10pt}
\caption{$T$-dependence of the Polyakov-loop at $\mu_{\rm B}=\mu_{\rm I}=0$. The LQCD date is taken from Ref.~\cite{Borsanyi_order}. }
\label{Fig_polyakov loop}
\end{figure}

\subsection{Chiral condensate}

In Fig.~\ref{chiral condensate}, we show the renormalized chiral condensate, the order parameter of chiral transition, at $\mu_{\rm B}=\mu_{\rm I}=0$. At high $T$, the renormalized chiral condensate becomes negative and deviates from the one in LQCD simulation. 
To overcome this difficulty, we  introduce the temperature dependence of $\frac{\partial M_{{\rm B},i}}{\partial m_f}$ and $\frac{\partial M_{{\rm M},j}}{\partial m_f}$ in Eq.~(\ref{sigma_f_h}). 
We assume the temperature dependence of $\frac{\partial M_{{\rm B},i}}{\partial m_f}$ and $\frac{\partial M_{{\rm M},j}}{\partial m_f}$ as follows.
\begin{eqnarray}
\frac{\partial M_{{\rm B},i}}{\partial m_f} = C_f^{{\rm B},i}g(T), 
\label{CgB_T}
\\
\frac{\partial M_{{\rm M},j}}{\partial m_f} = C_f^{{\rm M},j}g(T), 
\label{CgM_T}
\end{eqnarray}
where $g(T)$ is the function of temperature and takes value between 0 and 1. 
It is expected that $\frac{\partial M_{{\rm B},i}}{\partial m_f}$ and $\frac{\partial M_{{\rm M},j}}{\partial m_f}$ 
go to zero at high temperature, since hadrons disappear at high temperature. 
The explicit form of $g(T)$ which we use is given by following equation. 
\begin{eqnarray}
g(T) = \frac{1}{2}\left\{1+\tanh{((b_{\rm M} - T)e^{\left(\frac{c_{\rm M}}{T}\right)^{d_{\rm M}}}/a_{\rm M})}\right\}. 
\label{function of g}
\end{eqnarray}
We search parameters which reproduce the LQCD result of the normalized chiral condensate well.  
The obtained parameters are tabulated in Table~\ref{parameters of g} and the $T$-dependence of $g(T)$ is shown in Fig.~\ref{dMdm}.  
In Fig.~\ref{order parameter}, 
we also show the renormalized chiral condensate as a function of $T$, when the $T$-dependence of $\frac{\partial M_{{\rm B},i}}{\partial m_f}$ and $\frac{\partial M_{{\rm M},j}}{\partial m_f}$ is taken into account, as well as the Polyakov-loop. 
It is difficult to calculate to chiral susceptibility in our hybrid model. 
Hence, we calculate the derivative of chiral condensate with respect to $T$ and define the transition temperature $T_c^{C}$ where the absolute value of the derivative has a maximum. 
Obtained $T_c^{C}$ is 160MeV which is consistent with the chiral transition temperature $T_c^{C,{\rm L}}=154\pm 6$MeV found in LQCD simulations~\cite{{Borsanyi_Tc},{YAoki_Tc^(Z_3)}}. 

\begin{figure}[h]
\begin{center}
\vspace{0.5cm}
\includegraphics[width=0.35\textwidth]{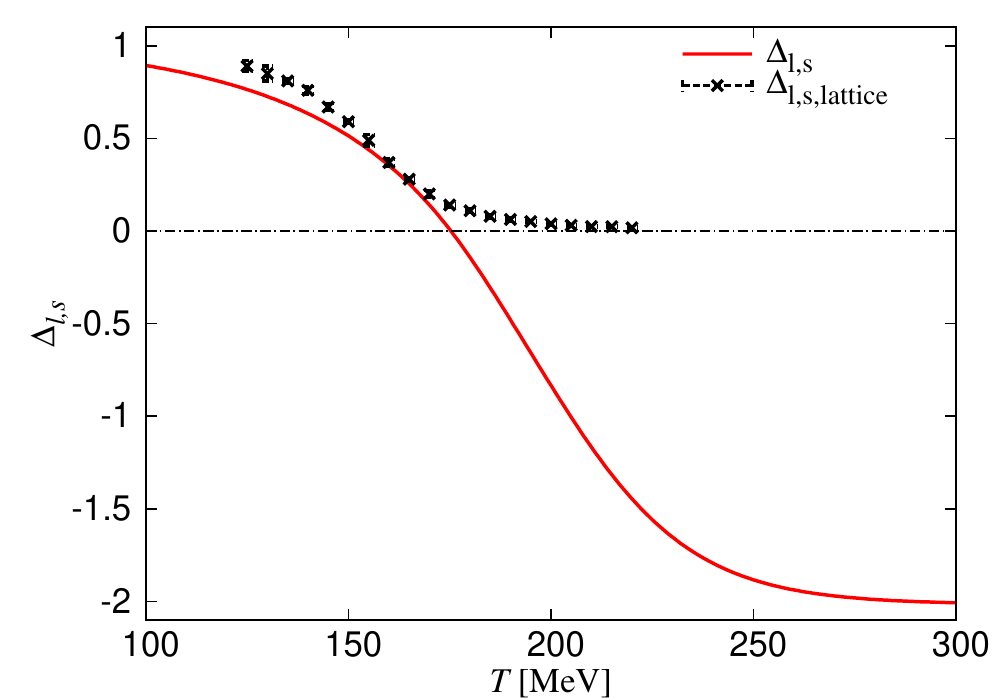}
\end{center}
\vspace{10pt}
\caption{$T$-dependence of the chiral condensate at $\mu_{\rm B}=\mu_{\rm I}=0$.The LQCD date is taken from Ref.~\cite{Borsanyi_order}. }
\label{chiral condensate}
\end{figure}

\begin{table}[h]
\centering
\begin{tabular}{lccr}
\hline
$\hspace{10mm}a_{\rm M}$&$\hspace{5mm}b_{\rm M}$&$c_{\rm M}$&$d_{\rm M}\hspace{5mm}$
\\ \hline
66.6654{\rm [MeV]}&198.644{\rm [MeV]}&172.781{\rm [MeV]}&4.78989\hspace{5mm}
\\ \hline
\end{tabular}
\caption{Parameters of $g$.}
\label{parameters of g}
\end{table}
\begin{figure}[h]
\begin{center}
\vspace{0.5cm}
\includegraphics[width=0.35\textwidth]{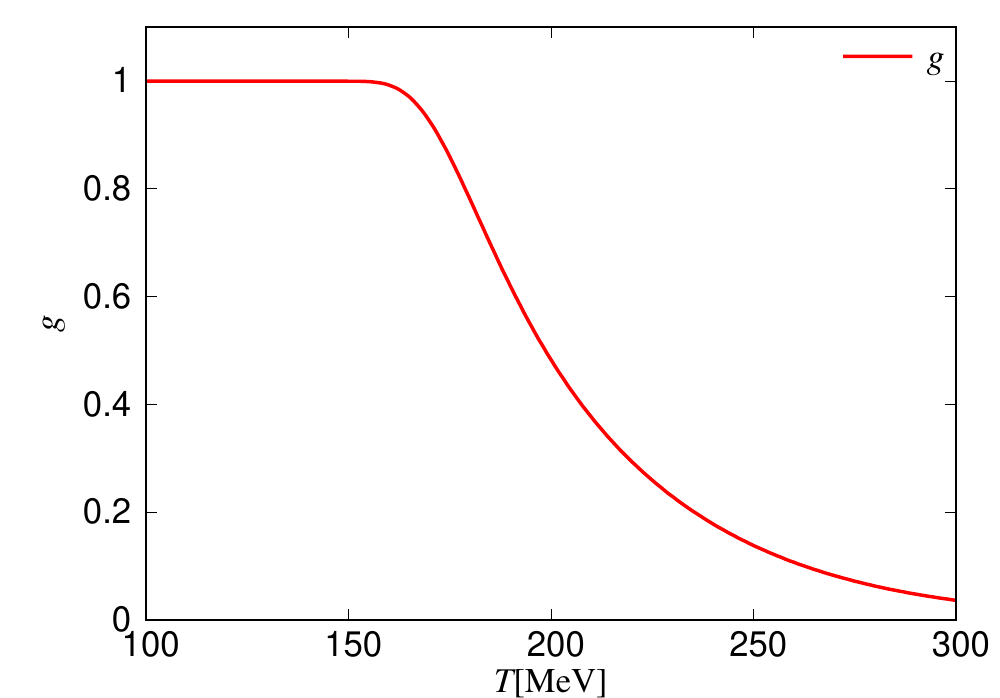}
\end{center}
\vspace{10pt}
\caption{$T$-dependence of $g(T)$. }
\label{dMdm}
\end{figure}

\begin{figure}[h]
\begin{center}
\vspace{0.5cm}
\includegraphics[width=0.35\textwidth]{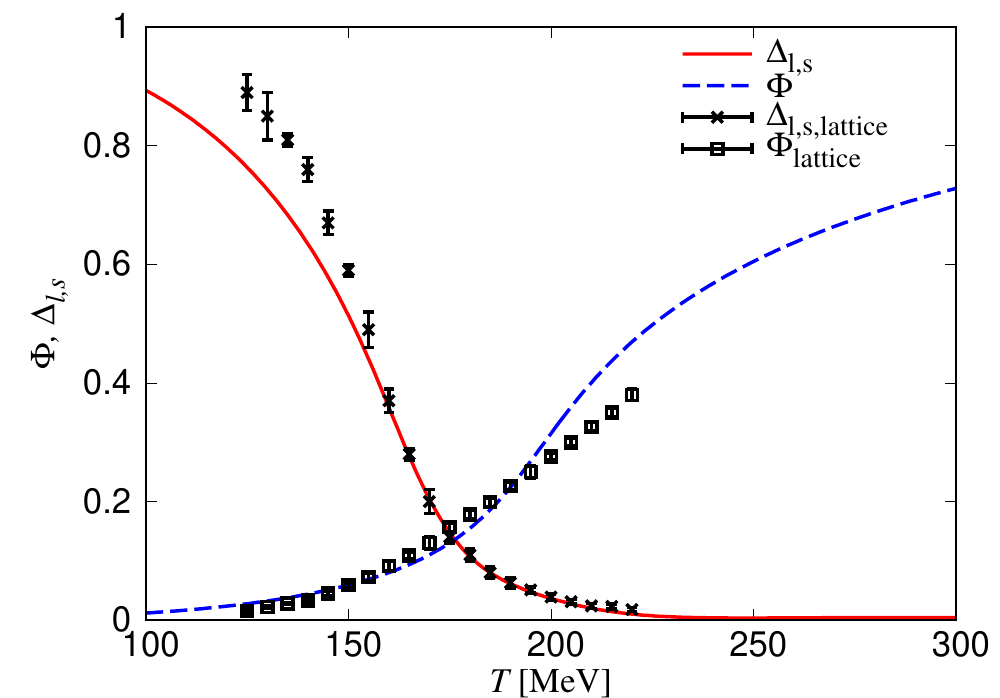}
\end{center}
\vspace{10pt}
\caption{$T$-dependence of the chiral condensate and Polyakov-loop at $\mu_{\rm B}=\mu_{\rm I}=0$. 
The result is obtained by using improved model with $T$ dependent ${\partial M_{{\rm B},i}\over{\partial m_f}}$ and ${\partial M_{{\rm M},j}\over{\partial m_f}}$. 
The LQCD date is taken from Ref.~\cite{Borsanyi_order}. }
\label{order parameter}
\end{figure}

\subsection{Transition temperature}

The transition temperatures obtained at $\mu_{\rm B}=\mu_{\rm I}=0$ are tabulated in Table~\ref{Tc_summary} as well as the ones obtained in LQCD simulations~\cite{{Borsanyi_Tc},{YAoki_Tc^(Z_3)}}. 
Table~\ref{Tc_summary} show that $T_{\rm c}^{\rm (s)}$ is obviously larger than $T_{\rm c}^{C,L}$ and $T_{\rm c}^{Z_3,L}$.

\begin{table}[h]
\centering
\begin{tabular}{lccr}
\hline
$\hspace{5mm}T_{\rm c}^{\rm (s)}$&$T_c^{\rm (P)}$&$T_c^{C,L}$&$T_c^{Z_3,L}\hspace{5mm}$\\ \hline
215{\rm [MeV]}&249{\rm [MeV]}&154$\pm$6[MeV]&170$\pm$7[MeV]\\
\hline
\end{tabular}
\caption{The summary table of transition temperatures $T_{\rm c}^{(s)}$ from entropy density and $T_{\rm c}^{(P)}$ from pressure in our hybrid model and chiral and $Z_3$ transition temperature $T_{\rm c}^{C,L}$, $T_{\rm c}^{Z_3,L}$ in LQCD calculation. $T_{\rm c}^{C,L}$ and $T_{\rm c}^{Z_3,L}$ is quoted by Ref.~\cite{{Borsanyi_Tc},{YAoki_Tc^(Z_3)}}.}
\label{Tc_summary}
\end{table}

\subsection{Thermodynamical quantity at finite chemical potential}

Next, we consider the $\mu_{\rm B}$ and $\mu_{\rm I}$ dependence of the hadron volume fraction function 
$f_{\rm H}$ at finite $\mu_{\rm B}$ and $\mu_{\rm I}$. 
The Taylor expansion of $f_{\rm H}(T,\mu_{\rm B}^2, \mu_{\rm I}^2)$ at $\mu_{\rm B}=\mu_{\rm I}=0$ is given by 
\begin{eqnarray}
f_{\rm H}(T,\mu_{\rm B}^2,\mu_{\rm I}^2) = f_{\rm H}(T,0,0) 
+ {\partial f_{\rm H}\over{\partial \nu_{\rm B}}}(T,0,0)\nu_{\rm B}\nonumber\\
+ {\partial f_{\rm H}\over{\partial \nu_{\rm I}}}(T,0,0)\nu_{\rm I}+\cdots . 
\label{expansion of fH}
\end{eqnarray}
Using the approximation up to the first order of $\nu_{\rm B}$ and $\nu_{\rm I}$ in the Taylor expansion 
(\ref{expansion of fH}), we can calculate the thermodynamical quantities when $\nu_{\rm B}=\mu_{\rm B}^2$ and/or $\nu_{\rm I}=\mu_{\rm I}^2$ are finite but not so large. 
For example, in Fig.~\ref{Fig_pressure at muB=300}, we show the $T$-dependence of pressure $P$ at $\mu_{\rm B}=300$MeV and $\mu_{\rm I}=0$.  
Our simple hybrid model reproduces the result of LQCD very well. 
The transition temperature $T_c^{(P)}$ is reduced to 236MeV at $\mu_{\rm B}=$300MeV.  
In Fig.~\ref{Fig_pressure at muB=300}, we also show the $T$-dependence of the hadron volume fraction function $f_{\rm H}$ at $\mu_{\rm B}=300$MeV and $\mu_{\rm I}=0$.   
At finite density, the hadron is suppressed at lower temperature than in the case of zero baryon density.  
This reduces the transition temperature $T_c^{(P)}$ at finite density. 

\begin{figure}[h]
\begin{center}
\vspace{0.5cm}
\includegraphics[width=0.35\textwidth]{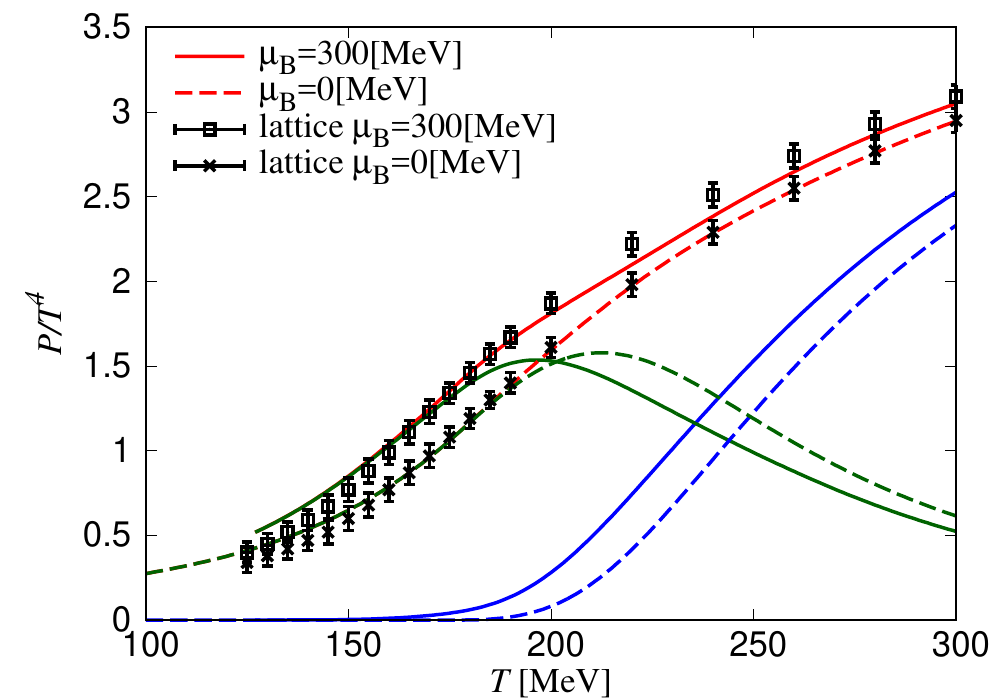}
\end{center}
\vspace{10pt}
\caption{$T$-dependence of the pressure at $\mu_{\rm B}=300{\rm MeV}$ and $\mu_{\rm I}=0$. 
The LQCD date is taken from Ref.~\cite{Borsanyi_entropy}. }
\label{Fig_pressure at muB=300}
\end{figure}

\begin{figure}[h]
\begin{center}
\vspace{0.5cm}
\includegraphics[width=0.35\textwidth]{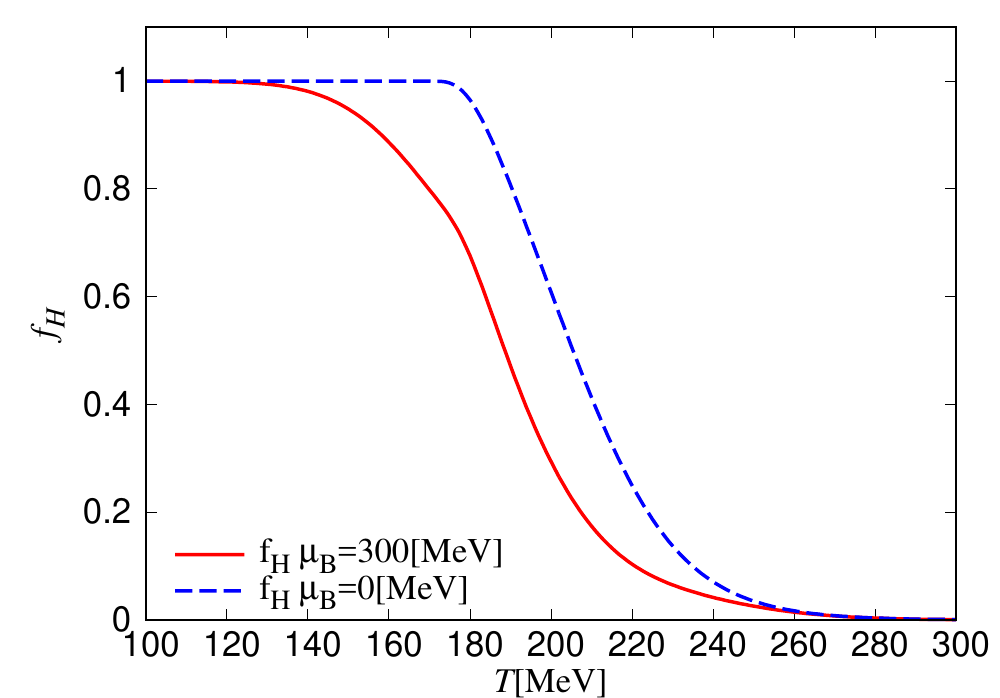}
\end{center}
\vspace{10pt}
\caption{$T$-dependence of the $f_{\rm H}$ at $\mu_{\rm B}=300{\rm MeV}$ and $\mu_{\rm I}=0$. }
\label{Fig_fH0 at muB=300}
\end{figure}

\section{Summary}
\label{summary}

In this paper, from the LQCD data, we have determined the ratio of hadron and quark contributions of thermodynamic quantities by using simple quark-hadron hybrid model. 
We have determined the transition temperature $T_{\rm c}^{\rm (s)}$ from the ratio of hadron and quark contribution of the entropy density.  
Our simple hybrid model can reproduce roughly chiral condensate and Polyakov-loop at the same time, but 
$T_{\rm c}^{\rm (s)}$ is obviously larger than chiral transition temperature $T_c^{C,{\rm L}}$ and $Z_3$ transition temperature $T_c^{Z_3,{\rm L}}$ in LQCD.  

The difference between $T_{\rm c}^{\rm (s)}$ and $T_c^{C,{\rm L}}$ can be understood as follows. 
In usual, it is natural that the temperature and/or density gives an effect opposite to the vacuum for the physical quantity.  
Hence, there is a tendency that the absolute value of chiral condensate or constituent quark mass decreases as the temperature and/or density increases, even in the theory without chiral symmetry.  
In fact, in the relativistic mean field theory of the quantum hadron dynamic without chiral symmetry, there is a tendency that the effective nucleon mass decreases as the density increases~\cite{SW}.    
It is also well-known that the QCD sum rule at finite density indicates the partial restoration of chiral symmetry in the normal nuclear matter~\cite{HL}. 
Recently, it was also shown~\cite{Makiyama} that, in the LQCD simulations of the two-color QCD, the hadron effect is very important 
in reducing the absolute value of chiral condensate at finite temperature and finite density, when the system is in confined phase. 
Hence, it can be considered that the chiral condensate decreases even in the hadron phase, when the temperature increases. 
This makes $T_c^{C,{\rm L}}$ lower than the confinement-deconfinement transition temperature $T_c^{(s)}$.  

On the other hand, a reason for the difference between $T_{\rm c}^{\rm (s)}$ and $T_c^{Z_3,{\rm L}}$ is rather unclear. 
However, this may indicates simply that $Z_3$-symmetry is not relevant symmetry for the confinement-deconfinement transition at finite temperature and the Polyakov-loop is not a good order parameter for the transition. 
Very recently, it was pointed out~\cite{Bazavov_Phi} that, in the 2+1 flavor LQCD simulations,  the temperature where a static quark entropy density (which is related to the Polyakov-loop) has a peak is close to the chiral transition temperature. 
Further study is needed in this direction. 

It is also interesting that $T_{1/2}^{f_{\rm H}}$ is close to the temperature where the interaction measure has a maximum. 
The study of the relation between the volume fraction function and the trace anomaly may be one of the interesting problems in future.

\noindent
\begin{acknowledgments}
The authors thank Atsushi Nakamura, Kouji Kashiwa, Junichi Takahashi, Masahiro Ishii, Junpei Sugano, Shuichi Togawa and Takehiro  Hirakida for useful discussions. 
M. Y. and H. K. are supported
by Grant-in-Aid for Scientific Research (No. 26400278 and No. 26400279) from the Japan Society for the Promotion of Science (JSPS). 

\end{acknowledgments}


\end{document}